\documentclass[10pt,conference]{IEEEtran}
\usepackage{cite}
\usepackage{amsmath,amssymb,amsfonts}
\usepackage[linesnumbered, ruled]{algorithm2e}
\usepackage{graphicx}
\usepackage{textcomp}
\usepackage{xcolor}
\usepackage{comment}
\usepackage[hyphens]{url}
\usepackage{fancyhdr}
\usepackage{hyperref}

\usepackage{grumble}

\usepackage[inline]{enumitem}
\setlist{noitemsep,topsep=0pt,parsep=0pt,partopsep=0pt}

\usepackage{setspace}
\usepackage[margin=5pt,font={stretch=0.9}]{caption}
\newcommand{\myparagraph}[1]{\noindent{\bf {#1}.}}
\newcommand{\projectname}{Zero-G}

\newcommand*\circled[1]{\tikz[baseline=(char.base)]{
            \node[shape=circle,draw,inner sep=1pt] (char) {#1};}}

\begin{document}
    \title{Zero\-G: A Pre-Decoder-Aware Decoder for Quantum Error Correction}

    \author{
        \IEEEauthorblockN{Peter Wegmann, Theofilos Augoustis, Aleksandra Świerkowska, Emmanouil Giortamis, and Pramod Bhatotia}
        \IEEEauthorblockA{Technical University of Munich, Munich, Germany \\
        \{peter.wegmann, theofilos.augoustis\}@tum.de}
    }
    
    \maketitle

\begin{abstract}
    Fault-tolerant quantum computing requires classical decoders that keep pace with the underlying hardware, translating syndrome measurements into corrections fast enough to avoid an exponential backlog. To meet this real-time constraint, {\em pre-decoders} have emerged as part of a hierarchical decoding approach to resolve simple, local errors before passing a sparser residual syndrome to a {\em strong decoder}. While pre-decoding should, in theory, speed up the strong decoder, in practice, the speedup is only marginal, since existing strong decoders are designed to decode dense syndromes and cannot exploit the sparsity provided by pre-decoders.

To address this, we present \projectname{}, a strong decoder designed for use alongside pre-decoders. As a stochastic approximate minimum-weight perfect
matching (MWPM) decoder, \projectname{} exploits sparse residual syndromes, dynamically trading latency for accuracy rather than relying on an all-or-nothing runtime-accuracy trade-off. By decoupling hardware control from the decoding core itself, we enable heterogeneous deployment across both FPGAs and CPUs without maintaining separate implementations. \projectname{} achieves a $10\times$ latency improvement over existing strong decoders at matching accuracy, with worst-case sub-$350\,\text{ns}$ decoding at code distances up to $d=15$, while scaling to 640 logical qubits on a single 128-core CPU and 32 logical qubits on a single AMD Versal V80 FPGA.

\end{abstract}

\section{Introduction}
    %- Context and motivation.
    %- The challenge in building fast decoders.
    %- Decoders are oblivious to pre-decoding.
    %- Research question.
    %- Research challenges.
    %- Our approach.
    %- Implementation.
    %- Evaluation.
    %- Contributions.

    \myparagraph{Context and motivation}
    Achieving fault-tolerant quantum computing (FTQC) requires quantum error correction (QEC) \cite{surface_code_fowler, backlog}, which operates in a continuous measure-decode-correct loop: syndromes are measured every code cycle, a classical \textit{decoder} infers the underlying errors, and a corresponding correction is applied before the next cycle begins. Because the decoder is the sole entity in this loop responsible for detecting and correcting errors, it sits squarely on the critical path, and its performance therefore directly gates that of the entire FTQC stack \cite{decoding_system, decoding_latency_influence}.

    This performance is characterized by two metrics: \textit{accuracy}, which directly determines the logical error rate (LER) \cite{threshold}, and \textit{latency}, the time required to produce a decoding solution. Latency directly determines whether the correction loop can keep running at all: since syndromes keep arriving every cycle regardless of whether the previous one has been decoded, a decoder that falls behind accumulates an ever-growing queue of undecoded syndromes, a \textit{backlog} that grows without bound and stalls the computation entirely, irrespective of how accurate the decoder is \cite{backlog}. This makes latency especially critical for superconducting architectures, whose QEC cycle lasts only $1\,\mu s$ \cite{helios, microblossom}, leaving decoders little room to keep pace before the next round of syndromes arrives. Real-time decoding is therefore a hard requirement for FTQC: without it, a computation cannot complete \cite{decoding_latency_influence}.

            \begin{figure}[t!]
    \centering
    \includegraphics[width=.95\linewidth]{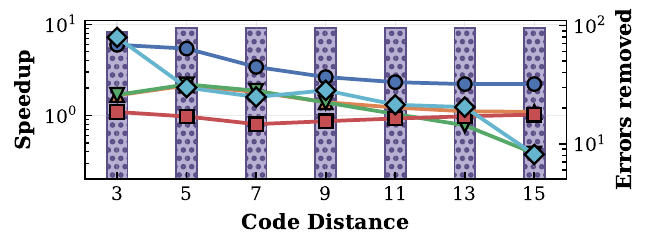}

    \vspace{-.5em}
    \includegraphics[width=1\linewidth]{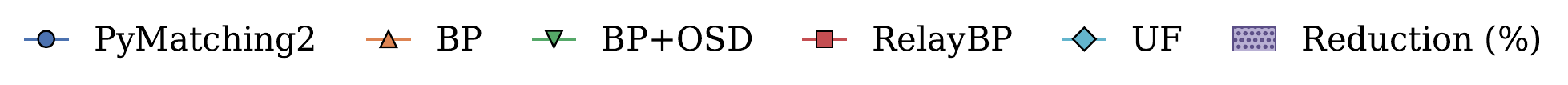}
    
    \vspace{-.5em}
    \caption{
        Motivation for \projectname{}, a decoder designed to be used with pre-decoders. \emph{While pre-decoders consistently filter out $\ge 90\%$ of errors, state-of-the-art decoders achieve only up to $6\times$ speedup, with diminishing returns at high code distances.}
    }
    \label{fig:motivation}
    \vspace{1pt}
    
\end{figure}

    % The challenge in building fast decoders
    \myparagraph{Decoding challenges}
    Meeting this real-time requirement is difficult because decoding itself becomes markedly more expensive as code distance grows. For the surface code \cite{surface_code, surface_code_fowler}, the most mature and widely studied QEC code, decoding reduces to Minimum-Weight Perfect Matching (MWPM). Exact MWPM solvers, however, offer no bound on worst-case latency and grow substantially more expensive with problem size, so they either cannot run within the decoding budget or require unrealistic hardware resources that do not fit within commodity accelerators when scaled to the code distances FTQC will require \cite{pymatching, helios, elastic_decoders}. Two strategies have emerged to close this gap. \textit{Approximate decoders} reduce the decoding solution space to lower latency, thus trading LER for latency \cite{union_finding, union_finding_qldpc, helios, actis, lcd_decoder, dst_decoder, riverlane_asic_decoder}. \textit{Pre-decoders} instead intercept the syndrome stream before it reaches a strong decoder, resolving simple, local error patterns quickly so that a secondary decoder needs only to resolve the remaining complex errors \cite{promatch, nvidia_ising, smith_predecoding, delfosse_hierarchical_decoding, Chamberland_predecoding}.

    \myparagraph{Neither approach is sufficient}
    Unlike approximate decoders, pre-decoders can improve LER without sacrificing latency, but only if paired with a fast secondary decoder \cite{nvidia_ising}: a slow secondary decoder cancels the pre-decoder's speedup, leaving the pipeline latency-bound once again. Both strategies therefore reduce to the same choice: an approximate decoder sacrifices LER for latency, while a pre-decoder pipeline built on a slow secondary decoder sacrifices latency for LER. Either choice ends up hurting LER, since a slow decoder leaves physical qubits idle for longer, and idle qubits accumulate additional errors through leakage and decoherence. Neither strategy alone, therefore, resolves the latency-accuracy trade-off that FTQC demands.

                \begin{figure*}[t]
    \centering
    \includegraphics[width=0.85\linewidth]{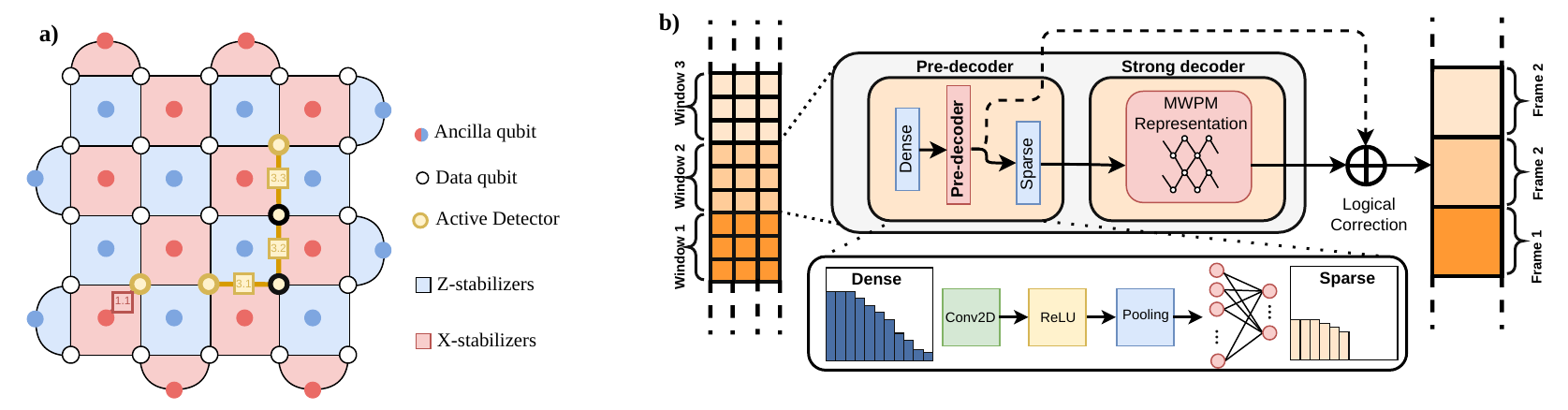}

    \vspace{-.5em}
    \caption{\textbf{(a)} Surface code memory and \textbf{(b)} sliding-window decoding with a hierarchical neural-network pre- and an MWPM strong decoder.}

    \label{fig:background}
    \vspace{-1.5em}
\end{figure*}

    % Decoders are oblivious to pre-decoding
    \myparagraph{Decoders are oblivious to pre-decoding}
    The reason this trade-off persists is structural: in practice, pre-decoders and strong decoders are deployed as two independent, mutually unaware stages \cite{delfosse_hierarchical_decoding, latte_predecoding_decoding_architecture, smith_predecoding}. Prior work has shown that pre-decoders resolve the large majority of shots, leaving only a sparse residual syndrome for the strong decoder \cite{nvidia_ising, promatch, smith_predecoding, clique, pinball, delfosse_hierarchical_decoding}. Yet existing strong decoders are built to decode dense syndromes: they operate on the \textit{full} matching graph induced by every detector in the code, regardless of how many of those detectors are actually flipped. Their runtime therefore scales with the number of detectors rather than with the number of residual errors, so even after a pre-decoder removes the bulk of the errors, the strong decoder pays the same asymptotic cost as if it were decoding the original, dense syndrome. In other words, the pre-decoder's work is structurally invisible to the strong decoder that follows it.

    We quantify this mismatch in Figure~\ref{fig:motivation}. While pre-decoders consistently filter out at least $90\%$ of errors across the code distances we study, state-of-the-art strong decoders translate this into no more than a $6\times$ latency speedup, with the gain shrinking further at higher code distances, precisely the regime FTQC needs to scale to. Closing this gap is not simply a matter of running an existing strong decoder more often or more cheaply: because current strong decoders operate solely on graph topology, a sparse syndrome and a dense one cost the same asymptotic work. Existing approximate decoders reduce latency by searching local regions instead of the full graph, but they do not concentrate on the residual, harder-to-match errors that pre-decoders leave behind \cite{promatch, pinball, clique, delfosse_hierarchical_decoding}, and hierarchical decoding schemes accelerate \textit{when} a strong decoder is invoked, not \textit{how} it decodes once invoked. To date, no strong decoder has been proposed that is co-designed with a pre-decoder in mind, explicitly exploiting sparse residual syndromes to reduce decoding latency.

    % Research question
    \statementLeftAccent{Research Question}{blue}{Can we design a \textit{pre-decoder aware} strong decoder that exploits sparse syndromes to reduce decoding latency?}

    % Research challenges
    \myparagraph{Research challenges}
    Answering this question is not easy, as a practical, pre-decoder-aware strong decoder must satisfy three challenges simultaneously. \textit{First}, the decoding algorithm must be fast enough to converge in tens to hundreds of nanoseconds, while avoiding the local optima that greedy matching heuristics are prone to getting stuck in. \textit{Second}, the decoder must be scalable both in code distance and in the number of decoding windows, to support sliding- or parallel-window decoding, and configurable, to let designers trade runtime for accuracy. \textit{Third}, to leverage the heterogeneous control architecture of quantum computers, the resulting decoder must cleanly map onto both CPU and FPGA execution.

    % Our approach
    \myparagraph{Our approach}
    We present \projectname{}, a strong decoder designed from the ground up to be pre-decoder aware. \projectname{} addresses the three challenges above through three co-designed layers. To converge within a tight nanosecond budget without getting trapped in local optima (Challenge 1), \projectname{} replaces the exhaustive augmenting-path search that locally-optimal solvers rely on \cite{pymatching} with a decoding algorithm based on multi-start stochastic local search: multiple lightweight matching attempts explore the sparse residual decoding subgraph concurrently, and a lightweight escape mechanism lets a search that stalls in a local optimum jump to a better region of the matching space, converging on a near-optimal solution within a small, bounded iteration budget. To make this budget configurable and to scale across code distance and decoding windows (Challenge 2), \projectname{} decouples the decoding task into a \textit{Zero-G Manager} and a \textit{Zero-G Decoding Core}: the Manager acts as the control plane, precomputing decoding matrices at compile time and buffering, dispatching, and, when needed, windowing the residual syndrome stream at runtime, while the Decoding Core acts as the data plane, streaming syndromes through independent \textit{Zero-G Decoder} instances and exposing the iteration budget as a first-class, per-instance parameter. Finally, because every Zero-G Decoder instance runs the same four-stage decoding workflow, i.e., extracting active detectors into a decoding subgraph, computing an initial low-latency matching, iteratively refining it, and deriving a Pauli frame update, the same algorithmic specification maps cleanly onto both CPU threads and FPGA fabric (Challenge 3), letting \projectname{} move decoder instances between backends without changing the control plane.

    % Implementation
    \myparagraph{Implementation and evaluation}
    We implement \projectname{} in C++, with the decoding core compiled onto both CPU thread pools and FPGA fabric from a single algorithmic specification.
    We evaluate \projectname{} on the surface code up to $d=15$ for $p \in [10^{-3}, 10^{-2}]$, matching MWPM LER while achieving sub-$350\,\text{ns}$ decoding latency on sparse residual syndromes. \projectname{} is also resource-efficient, enabling single-core multiplexing and naturally mapping onto FPGAs, achieving an accuracy-latency trade-off that no prior decoder achieves.

    % Contributions
    \myparagraph{Key contributions}
    Our contributions are as follows:

    \begin{itemize}
        \item \textbf{Pre-decoder aware decoding:} We present a configurable decoding pipeline that couples pre-decoders and strong decoders, explicitly exploiting the sparse residual syndromes to reduce communication bandwidth and runtime (\S~\ref{sec:overview}).
        \item \textbf{Stochastic approximate MWPM:} We propose a new class of QEC decoders based on stochastic local search, to approximate MWPM without incurring the cost of exact solvers (\S~\ref{sec:design}).
        \item \textbf{Decoupled decoding system:} We design a decoding system that separates hardware control from the decoding core via a dedicated decoding manager, enabling \projectname{} to be deployed transparently across both CPUs and FPGAs (\S~\ref{sec:zerog}).
        %\item \textbf{Greedy, pre-decoder-aware decoding as a design principle:} We show that a greedy, stochastic decoding strategy, while impractical on dense, unfiltered syndromes, becomes both faster and more accurate than local optimal approaches once combined with a pre-decoder, establishing sparsity-awareness as a first-class design constraint for strong decoders (\S~\ref{sec:design}).
        %\item \textbf{Full-stack implementation:} A complete C++ and FPGA implementation of \projectname{} and its pre-decoder interfaces, mapping cleanly onto both CPU and FPGA execution. \S~\ref{sec:hardware}
    \end{itemize}

\section{Background and Motivation}\label{sec:background}

    \subsection{Quantum Error Correction (QEC)}\label{subsec:background_qec}
        To suppress inherent hardware noise, quantum error correction (QEC) encodes logical qubits across multiple physical qubits using, e.g., the surface code (Fig. \ref{fig:background} \textbf{a)}). To preserve encoded logical qubits, continuous parity measurements of detectors generate syndromes by comparing outcomes across successive measurement rounds, with a flipped detector indicating an error. This syndrome is then passed to a classical decoder to compute a solution used to communicate logical qubit state to a Pauli Frame for tracking. 
        
        %However, to prevent an exponential backlog of syndromes, decoding must achieve high accuracy and complete within a maximum budget $t_{\text{budget}}$, typically $1~\mu\text{s}$ for superconducting systems \cite{backlog}.

        \statementLeftAccent{Challenge 1}{orange}{To prevent an exponential backlog of syndromes, decoding must achieve high accuracy and latencies below $t_{\text{budget}} \approx 1~\mu\text{s}$ for superconducting systems \cite{backlog}.}

        Addressing this challenge is difficult due to the complex topological structure of errors, since quantum codes feature (i) \textit{degeneracy} and experience both (ii) \textit{bit-} and \textit{phase-flips} errors. Thus, distinct physical errors can produce the same error pattern, so inferring the exact physical error is generally not possible, and two independent decoding tasks for $X$- and $Z$-type errors must be performed. In addition, errors rarely occur in isolation: adjacent errors form chains that cancel intermediate detectors \cite{lilliput_decoder}, so only the chain's \textit{ends} trigger a detector, making it impossible to directly tell which detectors belong to which chain. While recent observations suggest that most errors remain local and that long error chains are exponentially unlikely \cite{pinball, arqade}, chains can additionally terminate at boundary nodes, leaving single unmatched detectors.

        \statementLeftAccent{Observation 1}{gray}{Degeneracy and long-distance error chains make local syndrome interpretation ambiguous. Exact decoding, therefore, requires global decoding across the full detector graph rather than local approximations.}

    \subsection{Decoding}
        The general approach for decoding the surface code is to solve the Minimum-Weight Perfect Matching (MWPM) problem, an efficient approximation to the otherwise NP-complete decoding problem \cite{general_decoding_np_hard1, general_decoding_np_hard2, general_decoding_np_hard3}. However, even efficient approximations struggle to meet the required runtime \cite{decoder_bench, delfosse_decoder_selection}, motivating the use of fast but suboptimal decoders that result in a fixed accuracy-runtime trade-off curve \cite{spanning_decoder, lcd_decoder, union_finding}, or non-deterministic runtimes \cite{bp, bp_osd, relaybp, beam_search, dst_decoder}. 

        \statementLeftAccent{Challenge 2}{orange}{Current MWPM decoders force an all-or-nothing trade-off between runtime and accuracy, resulting in increased latency or reduced accuracy.}

        Two approaches have recently targeted the latency issue. First, sliding- and parallel-window decoding parallelizes over space by splitting the syndrome stream into overlapping windows \cite{sliding_window, parallel_window} to reduce decoding latency at the cost of increased decoding resources \cite{elastic_decoders}. Second, existing fast solvers can be further optimized for latency \cite{helios} or split across CPU/FPGA implementations \cite{microblossom}, which generally results in hardware resources of up to one FPGA per logical qubit. %. However, these trade specialized FPGA resources for latency, requiring up to one FPGA per logical qubit at $d=15$—hardware costs that make scaling to the thousands of logical qubits required for FTQC impractical.

        \statementLeftAccent{Observation 2}{gray}{Current decoding solutions require prohibitive hardware resources to satisfy real-time latency constraints.}

    \begin{figure*}[t]
    \centering
    \includegraphics[width=0.98\linewidth]{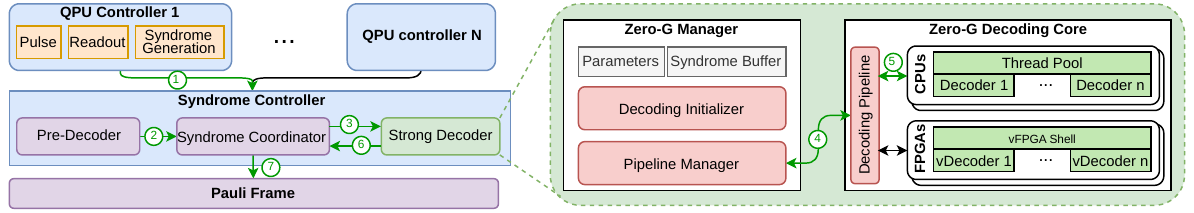}

     \caption{
        High-level overview of the decoding system, showing how \projectname{} integrates with existing decoding controllers within our pipeline.
        \emph{\protect\circled{\textup{1}} a dense syndrome is passed to the decoding controller, \protect\circled{\textup{2}} sparsified by a pre-decoder, \protect\circled{\textup{3}} passed to a syndrome coordinator, \protect\circled{\textup{4}} \& \protect\circled{\textup{5}} decoded using a CPU-based pipeline which \protect\circled{\textup{6}} returns the solution. Finally, \protect\circled{7} the result is passed to the Pauli frame.
    }
}
    \label{fig:system_overview}
    \vspace{-1.5em}
\end{figure*}

    \subsection{Pre-decoding}
        Since most syndromes exhibit short, locally confined error chains, fast heuristic algorithms can clear simple error configurations, forming the basis of hierarchical decoding which utilize a combined pre-decoder \cite{delfosse_hierarchical_decoding} and general decoder pipeline. Pre-decoders fall into two classes: syndrome-modifying (SM) \cite{promatch, nvidia_ising, bp_predecoding, Chamberland_predecoding, scalable_n_predecoding, smith_predecoding}, which sparsify a dense syndrome before passing the residual to a strong decoder, and non-syndrome-modifying (NSM) \cite{pinball, arqade, clique}, which attempt to solve the full syndrome and fall back to a strong decoder on failure. Our work targets SM pre-decoders that sparsify the syndrome in order to speedup the decoding of the secondary decoder.
        %Recent SM pre-decoders have achieved high error coverage but yield only average speedup for the secondary strong decoder \cite{promatch, nvidia_ising} (see Fig. \ref{fig:motivation}), revealing a gap between theoretical potential \cite{delfosse_hierarchical_decoding} and empirical performance \cite{nvidia_ising}.
        
        %\statementLeftAccent{Challenge 3}{orange}{Despite high error coverage, current pre-decoding pipelines yield only marginal speedups, as strong decoders are not designed to exploit the simplified workloads pre-decoders provide.}
        \statementLeftAccent{Challenge 3}{orange}{Despite achieving high error coverage \cite{promatch, nvidia_ising}, current pre-decoders speed up secondary general decoders only marginally (Fig. \ref{fig:motivation}), revealing a substantial gap between theoretical potential \cite{delfosse_hierarchical_decoding} and empirical performance \cite{nvidia_ising}.}
        
        Two bottlenecks limit this potential. First, exact MWPM solvers stay algorithmically complex regardless of input sparsity, since their worst-case runtime is dictated by problem size and not complexity. Second, approximate decoders like Union-Find (UF) offer better latency and FPGA tailored implementations, but their hardware utilization matches the problem's structure rather than the syndrome's complexity. Such implementations require up to one FPGA per logical qubit, resulting in severe resource underutilization for sparse syndromes \cite{helios, lcd_decoder}.

        %While utilizing fast decoders on the residual syndrome can theoretically reduce decoding latency, two primary architectural bottlenecks persist. First, exact MWPM solvers remain algorithmically complex regardless of input sparsity, since their worst-case execution time is governed by syndrome problem size, rather than problem complexity. While approximate decoders, such as Union-finding (UF) decoders, offer improved decoding latency and are well-suited for FPGAs, these implementations incur significant resource utilization to achieve this speedup. For instance, implementations like Helios \cite{helios} suffer from severe resource underutilization because their hardware footprint matches the decoding problem's structure, regardless of the syndrome's complexity. Thus, hardware utilization requirements of up to one FPGA per logical qubit fail to scale to the qubit count required by FTQC.
        
        \statementLeftAccent{Observation 3}{gray}{Strong decoders are decoupled from the pre-decoders, and are thus unable to exploit simplified syndromes to reduce runtime or hardware requirements.}

    \subsection{Problem Statement}
        Real-time decoding demands near-exact accuracy within a microsecond budget, yet strong decoders generally lock accuracy and latency into a fixed operating point and scale in runtime and hardware with code size, leaving them unable to exploit the sparse residual syndromes that pre-decoders produce. We therefore need a \emph{pre-decoder-aware} strong decoder whose effort scales with the \emph{complexity} of the residual syndrome. The resulting decoder must \emph{(i)} match the logical error rate of exact MWPM solvers, \emph{(ii)} expose a configurable accuracy-latency trade-off, and \emph{(iii)} bound each instance's hardware footprint to sustain many logical qubits per device.
% \newpage

\section{Overview}\label{sec:overview}

    We present \projectname{}, a strong decoder designed from the ground up to operate alongside a pre-decoder. Existing strong decoders treat pre-decoding as an opaque filtering step, whereas \projectname{} employs a decoding algorithm that exploits the sparsity of residual syndromes and exposes configuration parameters for different decoding constraints. To deploy the decoder across heterogeneous architectures, \projectname{} separates its control plane from its decoding data plane.

\begin{figure*}[t]
    \centering
    \includegraphics[width=0.95\linewidth]{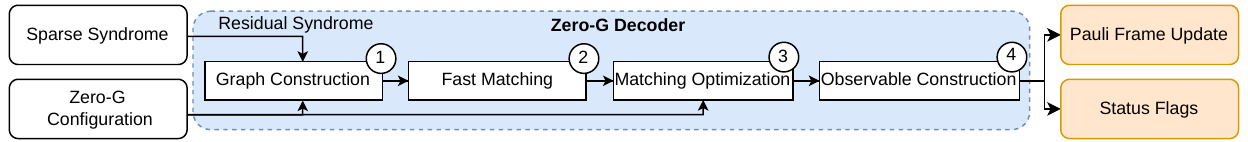}
    
    \caption{
        Workflow of the \projectname{} decoder.
        \emph{Given a sparse syndrome and decoding configuration, the \protect\circled{\textup{1}} decoding subgraph is constructed, from which \protect\circled{\textup{2}} a fast matching is constructed, \protect\circled{\textup{3}} optimized, and used to construct the \protect\circled{\textup{4}} logical observable. This is then returned as Pauli Frame Update and a decoding status.}
    }
    \label{fig:workflow}
    \vspace{-1.5em}
\end{figure*}

    Figure~\ref{fig:system_overview} \textbf{a)} locates \projectname{} within a full real-time decoding system and details its internal architecture
    (right). Figure~\ref{fig:workflow} shows the four-stage decoding workflow.% that \projectname{} executes on every residual syndrome.

    \subsection{System Components}
        We integrate \projectname{} into an experimentally validated full decoding system based on a pre-decoder and a strong decoder workflow \cite{decoding_system}. In the following, we present the five core components constituting the full decoding system (Figure~\ref{fig:system_overview} \textbf{a)}).
  
        \begin{itemize}
            \item \textbf{QPU controller:} Receives analog qubit signals given pulse level signals, and utilizes data acquisition modules for acquiring measurement signals. In addition, the QPU controller processes subsequent stabilizer measurement signals comprising the ancilla states to generate error syndromes.
            \item \textbf{Pre-decoder:} A dense, aggregated, syndrome stream is processed in order to remove locally-correctable errors and generate a residual sparse syndrome plus a residual Pauli frame update, leaving only complex error patterns.
            \item \textbf{Syndrome Coordinator:} Inspects the residual syndrome,  routes it to the strong decoder if further decoding is necessary, and finally combines the decoding solutions for communicating Pauli frame updates to the Pauli frame. The strong decoder is bypassed entirely when no residual errors remain.
            \item \textbf{Strong decoder (\projectname{}):} Decodes the residual syndrome received from the syndrome coordinator.
            \item \textbf{Pauli frame:} Receives Pauli frame updates from the syndrome coordinator and tracks the state of logical qubits.
        \end{itemize}

    \subsection{Stochastic Approximate MWPM}
        At the core of \projectname{} lies a single, hardware-agnostic decoding interface. Given a sparse residual syndrome, \projectname{} produces a decoding solution as a Pauli frame update, independent of whether the underlying decoder runs on a CPU or an FPGA. We realize this interface by separating \emph{control} from \emph{data}. The \textit{Zero-G Manager} serves as the control plane, managing decoder configurations, buffering incoming syndromes, and dispatching decoding tasks. The \textit{Zero-G Decoding Core} forms the data plane, streaming syndromes through \textit{Zero-G Decoder} instances deployed either on a CPU thread pool or on an FPGA. This separation lets \projectname{} move decoder instances between CPU and FPGA backends without modifying the control plane.

        %\peter{This is going to be section iv manager and decoder core.}

    \subsection{Workflow Walkthrough}
        We shortly present the full decoding workflow by tracing one syndrome through the full system in Figure~\ref{fig:system_overview}. \circled{1} \textit{QPU controllers} digitize readout pulses into per-round syndromes, aggregated into a dense syndrome stream by the decoding controller. \circled{2} The \textit{pre-decoder} removes locally-correctable errors, forwarding only a sparse residual syndrome and its corresponding residual Pauli frame update. \circled{3} The \textit{syndrome coordinator} checks whether the residual syndrome contains complex errors. If no errors are present, the strong decoder is directly bypassed, and \circled{7} the residual update is forwarded to the Pauli frame. Otherwise, a compressed residual syndrome is handed to \projectname{} (\S~\ref{subsec:fpga_architecture}).
        The \textit{Zero-G Manager} receives the incoming residual syndrome, and \circled{4} dispatches it to the \textit{Zero-G Decoding Core}, which \circled{5} uses a Zero-G decoder deployed either on a CPU or an FPGA. The decoder returns this update through the Zero-G Manager, together with status flags such as decode time, and \circled{6} passes the decoding solution to the syndrome coordinator. The pre-decoder's residual update is combined with the decoding solution, and \circled{7} then passed to the Pauli frame.

    \subsection{Design Challenges and Key Ideas}
        We develop and design \projectname{} around three key challenges. First, \textit{accuracy at scale}: to circumvent the backlog problem, a decoder must sustain the syndrome generation rate without sacrificing the logical error rate.
        Second, \textit{configurability}: different hardware and deployment systems require different LER/latency trade-offs, so \projectname{} exposes tunable decoding hyperparameters rather than a stale hard-coded decoding performance.
        Third, \textit{heterogeneous decoding system:} Realizing high-performance decoding logic typically forces a choice between the high sequential clock frequency of general-purpose CPUs and the spatial parallelism of FPGAs. The core engineering idea of \projectname{} is to decouple the core decoding logic from the decoding control and to use a decoding algorithm suitable for both CPU and FPGA deployment, so that a single algorithmic specification can be deployed on both CPUs and specialized FPGA hardware.

\section{\projectname{}}\label{sec:zerog}
    We present the realization of the core components of \projectname{}. Following the control/data separation introduced in \S~\ref{sec:overview}, the \textit{Zero-G Manager} implements the control plane and the \textit{Zero-G Decoding Core} implements the data plane, which streams decoding tasks through \textit{Zero-G Decoder} instances in a pipelined fashion.

    \subsection{The Zero-G Manager}\label{subsec:zerog_manager}
        The \textit{Zero-G Manager} is the control-plane component of \projectname{}, which ingests syndrome streams, collects decoding solutions, and manages decoder parameters. We split its responsibilities into a runtime component that manages the live data stream and holds decoder state, and a compile-time component that precomputes the decoding matrices used at runtime.

        Before any syndrome is decoded, the Zero-G Manager utilizes the \textit{Decoding Initializer} to precompute the decoding matrices and derive decoding parameters (\S~\ref{subsec:impl_init}). This includes selecting the decoder \textit{parameters} appropriate for the target LER/latency trade-off (\S~\ref{sec:overview}) and constructing the static routing matrices used to build decoding subgraphs at runtime. While this stage is resource-intensive at compile time, it is executed only once.

        During runtime, the manager maps and streams jobs from the Syndrome Buffer to the Zero-G Decoding Core. The \textit{Pipeline Manager} is agnostic to whether the underlying processing elements are CPU threads or deployed on FPGAs, which enables \projectname{} to target both backends from a single control plane. Beyond simple shot-by-shot dispatch, the Pipeline Manager can be extended to perform sliding- or parallel-window decoding \cite{sliding_window, parallel_window} by partitioning the incoming syndrome stream into decoding windows, scaling decoding throughput without exposing additional control logic to the Zero-G Decoding Core. Because the Manager reports per-shot decode times and accepts decoder configurations through its external interface, a higher-level runtime can supervise, reconfigure, and account for individual decoder instances without touching the data plane.
        
    \subsection{The Zero-G Decoding Core}\label{subsec:zerog_decoding_core}
        The \textit{Decoding Core} serves as the dedicated data-plane execution engine of \projectname{}. It receives a continuous stream of residual syndromes from the \textit{Pipeline Manager} and instantiates one or more \textit{Zero-G Decoder} instances, which can either be deployed on an FPGA or CPU threads.
        Since every decoding instance implements the same decoding algorithm (\S~\ref{sec:design}), a single algorithmic implementation can be compiled to either backend, fulfilling the heterogeneity goal from \S~\ref{sec:overview}.
        Upon completing a decoding task, an individual instance immediately fetches the next pending syndrome directly from the syndrome stream, decoupling decoding progress from solution reporting to the \textit{Zero-G Manager}.

\section{Zero-G Decoding Core}\label{sec:design}

    % General
    This section walks through the design of our decoding algorithm, a stochastic approximation to MWPM \cite{surface_code} consisting of four stages. We did not develop our algorithm as a single theory-driven phase, but rather through an iterative, empirically driven process, introducing each additional stage to improve upon the one before it.

    % Runtime complexity
    Our main algorithmic goal is to decouple decoding complexity from code distance $n$ and shift this onto the number of active detectors in a syndrome $n_d$. Global decoders such as union-find \cite{union_finding, helios} or sparse blossom \cite{pymatching} have empirical scaling of $\mathord{\sim}O(n)$ with respect to code distance \cite{pymatching}, but the actual information content of a syndrome under realistic, low physical error rates is much sparser than $n$. Our decoder instead scales with the number of \textit{active} detectors $n_d$, giving ROMA a worst-case runtime of
    \begin{equation}
        T(n_d) = \mathcal{O}\!\left(n_d^2 \left(\log n_d + k \cdot l\right)\right)
    \end{equation}
    where $k$ and $l$ are hyperparameters introduced below. Because the expected number of active detectors scales with the physical error rate rather than the code distance ($n_d \ll n$) \cite{delfosse_hierarchical_decoding, wootton_decoding}, our approach is very attractive for low-physical-noise regimes, in combination with pre-decoders \cite{promatch, nvidia_ising}.

    % Overview of the decoding pipeline
    Specifically, our pipeline consists of four distinct stages, as illustrated in Fig.~\ref{fig:workflow}. Given a syndrome, we (i) construct a detector subgraph from the fired detectors extracted from the syndrome (\S~\ref{subsec:impl_graph}). (ii) We compute an initial fast matching solution $M_{\text{fast}}$ using the \textit{GRDY} algorithm (\S~\ref{subsec:impl_grdy}). (iii) Starting from $M_{\text{fast}}$ we perform stochastic local search to find an improved matching $M_{\text{local}}$ (\S~\ref{subsec:impl_rama}). (iv) In the last step we refine $M_{\text{local}}$ via 2-augmentation paths into $M_{\text{2-opt}}$ (\S~\ref{subsec:impl_roma}). A pseudo-algorithm of the full algorithm is given in Algorithm~\ref{alg:greedy_mwpm}

\begin{figure}[t!]
    \centering
    \includegraphics[width=.99\linewidth]{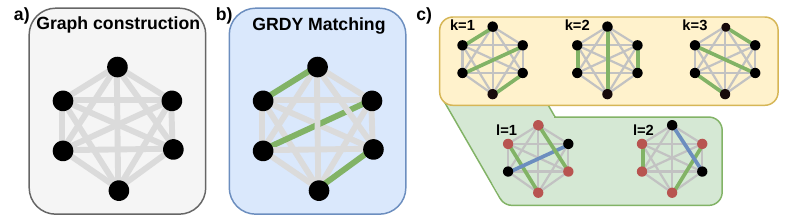}

    %\framebox{\begin{minipage}[c][5cm][c]{0.9\linewidth}
    %    \centering
    %    \textbf{\peter{Placeholder figure for our fpga integration/how everything connects}} \\
    %\end{minipage}}
    
    \vspace{-5pt}
    \caption{
        Visualization of graph construction and matching workflow. \emph{\textbf{a)} A complete graph is built from the syndrome, \textbf{b)} an initial matching computed, and \textbf{c)} refined using RAMA and ROMA.}
        }
    \label{fig:impl_example}
    \vspace{1pt}
    
\end{figure}

    \subsection{Initialization}\label{subsec:impl_init}
        Before deployment, an offline initialization phase builds two static lookup tables: a distance matrix and a path-mask matrix based on the code distance and the detector error model. The distance matrix stores all-pairs shortest-path distances between detectors and a boundary node, with edge weights mapped from physical error probabilities $p$ to $w = -\ln(p)$, differing slightly from PyMatching2's $\ln\!\left(\frac{1-p}{p}\right)$ weighting \cite{pymatching}. The path-mask matrix stores how the shortest path between matched detectors flips the logical qubit, turning the matching-to-observable into a table lookup rather than graph traversal.

    \subsection{Graph Construction}\label{subsec:impl_graph}
        Our syndrome decoding starts by constructing a fully connected detector subgraph from the set of active detectors in the sparse syndrome. Let $V_d$ be the set of $n_d$ active detectors, let $b$ be a single global boundary node, and $E$ be a list with an edge between every pair of active detectors. We build a fully connected detector subgraph $G = (V_d \cup \{b\}, E)$ with weights read directly from the pre-computed distance matrix. This step is shown as \textsc{ConstructGraph} in Alg.~\ref{alg:greedy_mwpm}.

 {
 \newcommand{\myfontsizealgorithms}{\fontsize{8}{8}\selectfont}
 \LinesNumbered
 \setlength{\parskip}{0pt}
 \begin{algorithm}[t]
     \myfontsizealgorithms
     \SetAlgoLined
     \SetKwProg{Fn}{Function}{:}{}
     \newcommand\mycommfont[1]{\fontsize{8}{8}\textcolor{codegreen}{#1}}
     \SetCommentSty{mycommfont}
     \SetNoFillComment
     \underline{{\bf decode}(sparse\_syndrome: S, decoder\_parameters: k, l)} \\
     \Begin{
        \tcc{Stage 1: Construct a fully connected detector subgraph}
        $E \gets$ \text{ConstructGraph}($S$)

        \tcc{Stage 2: Compute fast initial greedy matching}
        $M_{\text{fast}} \gets$ \text{FastGreedyMatching}($E$)

        \tcc{Stage 3: Perform global optimization using $k$-RAMA}
        $M_{\text{best}} \gets M_{\text{fast}}$
        
        \For{$i \gets 0$ \KwTo $\text{k}$}{
            $M_{\text{aug}} \gets M_{\text{fast}}$

            $E_{\text{rand}} \gets \text{Shuffle}(E)$

            \ForEach{$p \in E_{\text{rand}}$}{
                \tcp{Apply 2-augmentation path}
                $M_{\text{aug}} \gets M_{\text{aug}} \oplus \text{AugmentPath}(d=2, M_{\text{aug}}, E)$
            }
            \tcp{Keep the best matching}
            \uIf{$w(M_{\text{aug}}) < w(M_{\text{best}})$}{
                $M_{\text{best}} \gets M_{\text{aug}}$
            }
        }
        $M_{\text{local}} \gets M_{\text{best}}$

        \tcc{Stage 4: Perform local optimization using $l$-ROMA}
        $M_{\text{2-opt}} \gets$ $l$\text{-ROMA}($M_{\text{local}}, E, \text{l}$)

        \tcc{Convert the matching to logical observable}
        $\text{Observables} \gets$ \text{CalculateObservable}($M_{\text{2-opt}}$)

        \Return Observables
     }
     \caption{The \projectname{} decoding algorithm.
     }
    \label{alg:greedy_mwpm}
 \end{algorithm}
 }

    \subsection{Fast Solution}\label{subsec:impl_grdy}
        With the subgraph constructed, we attempt an initial matching using a greedy matching \cite{greedy_algorithm} by sorting the $O(n_d^2)$ edges by weight and commit each edge to $M_{\text{fast}}$ if both are still unmatched, shown as \textsc{FastGreedyMatching} in Alg.~\ref{alg:greedy_mwpm}. Although this stage is dominated by the sort's $\mathcal{O}(n_d^2 \log n_d)$ runtime, it remains fast enough to run on every shot. On its own, however, it leaves a substantial LER gap to the optimal MWPM solution, since early local matches block better global pairings later. We therefore treat $M_{\text{fast}}$ as the starting point for the local search that follows.
        
    \subsection{Global Solution Converging}\label{subsec:impl_rama}
        After constructing an initial matching, we want to improve it without relying on an expensive full global solver. Our first attempt is a single pass of the Random Augmentation Matching Algorithm (RAMA) \cite{RAMA_algorithm} over $M_{\text{fast}}$, which shuffles the edge set and greedily applies 2-augmenting-path improvements. While a single RAMA pass improves $M_{\text{fast}}$, this process is sensitive to the order of the path augmentations, since different orderings find different local minima.

        To escape local minima, we repeat edge shuffling $k$ times independently, keeping the best matching by total weight across all $k$ iterations ($k$-RAMA). Restarting from diverse edge orderings, rather than continuing to improve a single trajectory, allows us to explore different augmenting-path sequences and escape local minima we would otherwise not be able to. To keep the per-iteration cost of $k$-RAMA linear, we restrict augmenting paths to length 2, resulting in a runtime of $O(k \cdot n_d)$. A full algorithmic description is given in Alg.~\ref{alg:greedy_mwpm}. 
        \begin{figure*}[t!]
    \centering
    \includegraphics[width=.97\linewidth]{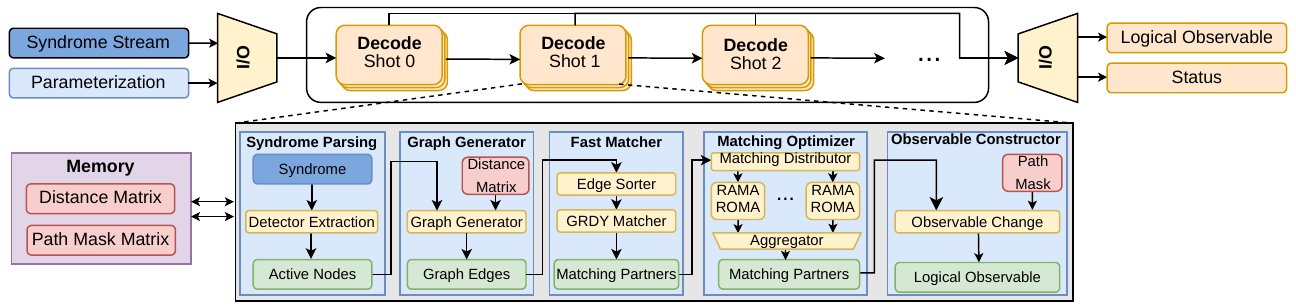}
    
    \vspace{-5pt}
    %\caption{
    %    Overview of the hardware implementation of the Zero-G Decoding Core integrated into a decoding pipeline, illustrating the processing of a continuous input syndrome stream and resulting Pauli frame updates as logical observables through I/O interfaces. \emph{A dedicated memory region stores precomputed decoding structures, such as the distance and path mask matrices. The architecture presents the decoding phases as sequential stages of detector extraction and observable construction, with a parallelized matching optimization stage that performs the optimization in parallel and finally outputs a logical observable.}
    %}
    \caption{
    Overview of the hardware implementation of Zero-G Decoding Core within a decoding pipeline, processing a continuous syndrome stream via dedicated I/O interfaces. \emph{A dedicated memory region stores precomputed decoding matrices utilized by the decoder.}
}
    \label{fig:hardware_design}
    \vspace{-1.5em}
    
\end{figure*}

    \subsection{Localized Solution Refinement}\label{subsec:impl_roma}
        Restarting from scratch at each iteration imposes an inherent ceiling, since we do not accumulate improvements across iterations the way a continuous refinement can. Rather than repeatedly resetting to $M_{\text{fast}}$, in the following, we allow matchings to accumulate augmentations across rounds, without discarding progress after each restart by utilizing the Random Order Augmentation Matching Algorithm (ROMA) \cite{ROMA_algorithm}. $l$-ROMA takes $M_{\text{local}}$ and, over $l$ rounds, shuffles the edge set and applies a 2-augmenting path to every edge in that order, continuously improving the same matching into $M_{\text{2-opt}}$. This step is shown as $l$-\textsc{ROMA} in Alg.~\ref{alg:greedy_mwpm}.

        By first identifying a local minimum through $k$-RAMA and then refining it via $l$-ROMA, our decoder closes the remaining LER gap to PyMatching2.

    \subsection{Adaptive Stage Tuning}\label{subsec:impl_configurability}
        A key advantage of our complementary stochastic stages is that the same pipeline can be tuned toward two different goals by splitting the decoding budget between $k$-RAMA restarts and $l$-ROMA rounds. When optimizing for latency, we use higher $k$ values to add $k$-RAMA rounds, as this leads to faster convergence. When optimizing for accuracy, $k$-RAMA is combined with additional $l$-ROMA rounds, resulting in further refinements when combined rather than relying on restarts alone.

\section{Hardware Design}\label{sec:hardware}
    While the CPU implementation of \projectname{} meets the round budget on average, its tail latency is governed by scheduling jitter and cache misses. Since real-time decoding must bound the worst case, we additionally implement the decoding stages of \S~\ref{sec:design} on an FPGA, where every operation completes in a fixed number of cycles and the latency tail depends only on the syndrome. Both backends implement identical decoding stages and remain interchangeable behind the Zero-G Manager.

    \subsection{Pipelined Architecture Design}
        The decoding algorithm of \projectname{} lends itself to a streaming pipeline because every stage performs bounded work over the active detectors of a single shot. The resulting design, shown in Figure~\ref{fig:hardware_design}, consumes a stream of per-shot syndromes and emits, per shot, one logical observable and a decoding status word. Independent shots flow through the stages back-to-back, so the pipeline sustains a continuous syndrome stream without stalling.

        \myparagraph{Stages}
            The pipeline implements the decoding stages of \S~\ref{sec:design} as dedicated hardware units, beginning with an ingest stage that parses the active-detector list of a shot and passes it to graph construction, which builds the edge list by obtaining the distance and path mask of every candidate edge and capturing both in a per-shot cache. This single pass covers every matrix entry the shot needs, so the remaining stages run on registered on-chip state and never wait on matrix memory. The matching stage pairs the active detectors by selecting the minimum-weight eligible edge through a comparator tree, producing the same matching as a stably sorted edge list without the data-dependent latency of sorting in hardware. Augmentation refines this matching over the configured number of depth-2 rounds, and observable construction folds the cached path masks of the matched pairs into the final logical observable.

        \myparagraph{Fast path}
            Residual syndromes concentrate at low detector counts, so the pipeline provides two implementations of these stages. A fast path serves shots with at most ten active detectors using fully partitioned scratch state, ten-entry match arrays, and four-bit node indices, while a generic path serves larger shots up to the configured capacity of 128 active detectors. Both paths implement identical algorithmic semantics, so the fast-path threshold only selects which implementation decodes a shot.

        \myparagraph{Randomization}
            The augmentation stage traverses the active detectors in a shuffled order to escape local minima (\S~\ref{subsec:impl_rama}). True random shuffling is ill-suited to hardware because it requires an entropy source and data-dependent memory permutations. Instead, the pipeline generates one permutation per restart with a hardware Fisher-Yates shuffle driven by a hash-based xorshift PRNG \cite{marsaglia_xorshift_2003}, seeded deterministically by the shot content and the restart index, and reuses it across the refinement rounds of that restart. %The resulting matchings can deviate slightly from the software reference, which draws its permutations from a system-level random source. We quantify this deviation in \S~\ref{sec:evaluation} and observe no measurable impact on the logical error rate.

        \myparagraph{Configurability}
            The number of augmentation iterations is a runtime parameter carried in each command header and bounded by a compile-time maximum, so a deployment can move along the accuracy-latency curve per batch without re-synthesizing the design. This realizes the accuracy-bounded and time-bounded operating modes of \S~\ref{subsec:impl_configurability} directly in hardware.

        \myparagraph{Replication}
            The multi-start search of \projectname{} restarts $k$ times from the same greedy matching, which makes the $k$ restarts data-independent. The pipeline exploits this independence by replicating the augmentation datapath $k$ times. All replicas read the shared per-shot distance cache, and a comparator tree selects the restart with the largest accumulated weight reduction, trading LUT resources for a single parallel augmentation pass instead of $k$ sequential ones. When a fixed $k$ is undesirable, the same restarts execute as a runtime loop over a single datapath instance at the cost of additional cycles per shot.

    \subsection{FPGA Architecture}\label{subsec:fpga_architecture}
\begin{figure}[t!]
    \centering
    \includegraphics[width=.98\linewidth]{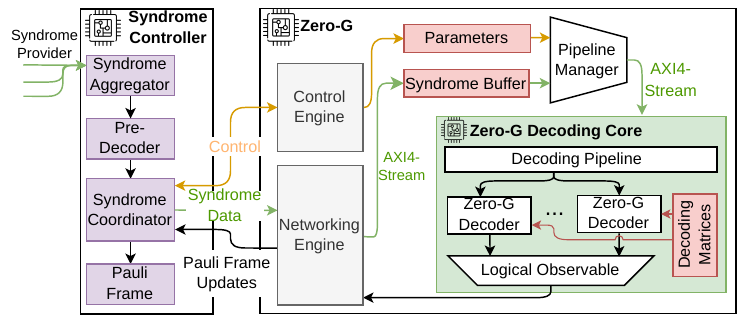}

    %\framebox{\begin{minipage}[c][5cm][c]{0.9\linewidth}
    %    \centering
    %    \textbf{\peter{Placeholder figure for our fpga integration/how everything connects}} \\
    %\end{minipage}}
    
    \vspace{.5em}
    \caption{Overview of the \projectname{} FPGA deployment. \emph{The syndrome controller exchanges syndromes, control signals, and Pauli frame updates with the Zero-G FPGA over AXI4.}}
    \label{fig:fpga}
    \vspace{1pt}
    
\end{figure}

        We deploy the pipeline within the split control architecture of today's quantum systems, shown in Figure~\ref{fig:fpga}. The \emph{syndrome controller} aggregates syndromes, pre-decodes them, coordinates residual syndromes, and tracks the Pauli frame alongside the QPU control logic, which leaves it few resources to spare for decoding. Since a real-time strong decoder exceeds that budget, control architectures provision dedicated decoding hardware, and the \emph{Zero-G FPGA} fills this role by hosting the strong-decoder pipeline described above, receiving from the syndrome controller only the sparse residual syndromes that survive pre-decoding.

        \myparagraph{Syndrome communication}
            The syndrome controller streams commands to the Zero-G FPGA, each beginning with a header word that carries the decoding parameters, including an opcode that distinguishes matrix loading from decoding, the input format, the batch size, and the restart and iteration counts. Since the syndrome controller inspects the residual syndrome to determine whether further decoding is necessary, active detectors can be simultaneously extracted. To reduce communication bandwidth, a major bottleneck \cite{afs_decoder, realtime_decoding_challenges, delfosse_hierarchical_decoding, smith_predecoding}, we therefore transmit only a compact active-detector list. Thus, a shot with at most ten active detectors fits in a single 512-bit beat, so interface cost scales with syndrome complexity rather than full size. For every decoded shot, the pipeline returns a single output word containing the packed observable mask, status flags, and the measured per-shot cycle count, keeping all control logic in the Zero-G Manager (\S~\ref{subsec:zerog_manager}).

        Per-shot state is small and stays on-chip in fully partitioned arrays. The distance and path-mask matrices instead grow quadratically with the detector count and dominate the memory footprint, so their placement determines both decoding latency and deployment density. \projectname{} therefore supports a \emph{latency-optimized} and a \emph{density-optimized} matrix placement.

        \myparagraph{Latency-optimized placement}
            The latency-optimized placement exploits that a shot with at most ten active detectors touches at most $45$ of the 5.6M detector-pair entries the $d=15$ matrices hold. The syndrome controller therefore projects the matrices onto the active set and streams the corresponding records with each shot, each packing the 32-bit distance and the 40-bit path mask of one detector pair, while the detector-to-boundary records that every shot needs remain resident in BRAM. The decoder's memory footprint thereby scales with residual sparsity, whereas keeping the full matrices resident on-chip would consume $73\%$ of the URAM of a Versal V80.

        \myparagraph{Density-optimized placement}
            The density-optimized placement stores the matrices in the high-bandwidth memory (HBM) of the accelerator card, shared read-only among all decoder instances, with up to eight outstanding reads in flight to hide access latency. It supports arbitrary code distances and dense multi-instance deployment at the cost of higher per-lookup latency. The decoding core addresses the HBM-resident matrices through virtual addresses, so instances share them without driver-side bookkeeping.

\section{Methodology}\label{sec:methodology}
    % How are the experiments etc. setup

    \subsection{Decoding Framework}
        \myparagraph{Experimental Setup}
            To evaluate and compare different decoders, we implement a custom evaluation framework in Python 3.14. This framework is used to benchmark modular decoding pipelines, including standalone strong decoders and configurations with pre-decoders. We integrate our \projectname{} C++ implementation using nanobind \cite{nanobind}. To collect decoding statistics, we conduct Monte Carlo simulations of surface code memory circuits and pre-generate error syndromes via Stim \cite{stim_gidney}. In our decoding pipeline, syndromes are initially processed by a pre-decoder, which returns a partial correction and a residual syndrome, which is subsequently forwarded to the strong decoder and lastly combined. Unlike several prior works that decode either $X$ or $Z$ errors \cite{promatch, helios, clique, astrea}, our decoder corrects both error types simultaneously \cite{pymatching}.

        % Metrics
        \myparagraph{Metrics}
            We evaluate the performance of \projectname{} using three primary metrics: \circled{1} Logical Error Rate (LER), \circled{2} average decoding latency, and \circled{3} decoding latency distribution.
            The LER is defined as the fraction of decoding solutions in which the logical observables differ from the initial logical state. We calculate this metric by aggregating up to $10^8$ Monte Carlo shots.
            To evaluate the worst-case decoding latency while filtering out system-level jitter (e.g., CPU context switches) unrelated to decoding latency, we track single-shot execution times and report 99th-percentile latencies \cite{decoding_latency_influence}.

        \myparagraph{Baselines}
            We choose (uncorrelated) PyMatching 2 \cite{pymatching} as our primary baseline for its high accuracy, high-performance implementation, and predictable runtime. BP-OSD \cite{bp_osd} decoders offer competitive thresholds at slightly higher latency, but their runtime is non-deterministic on standard topological surface codes, making PyMatching 2 the industry standard for comparison. We additionally evaluate the speedup \projectname{} achieves over Union-Find, PyMatching, BP-OSD, BP, and RelayBP in Fig.~\ref{fig:motivation}.

    \subsection{Circuits}
        We utilize rotated surface code memory circuits generated using the NVIDIA CUDA-Q QEC library v0.1.2 \cite{nvidia_ising}, which interfaces with Stim v1.16.0 \cite{stim_gidney}. We apply the SI1000 circuit-level noise model \cite{si1000_error_model} and sweep physical error rates in the range $p \in [10^{-4}, 10^{-2}]$ to evaluate performance across realistic hardware configurations. To evaluate scalability, we vary the code distance $d \in [3, \dots, 21]$ and fix the number of syndrome extraction rounds to $d + 1$. This range captures the medium-distance regime where topological boundaries actively interact with error chains, beyond which exponential suppression of logical errors tends to absorb local decoding mispredictions \cite{surface_code_fowler}.

    \subsection{Pre-Decoder}
        To evaluate the performance of \projectname{}, we select three available pre-decoders: Smith, Promatch, and Nvidia-Ising \cite{smith_predecoding, promatch, nvidia_ising}. 
        For Nvidia-Ising, we use the pre-trained \textit{fast} model deployed on a single NVIDIA H200 GPU \cite{nvidia_ising}, trained on syndrome data simulated under a modified SI1000 circuit-level noise model at a fixed physical error rate of $p_{\text{training}} = 5 \times 10^{-3}$.

    \subsection{Hardware}
        We implement \projectname{} in C++17 and provide Python bindings utilizing nanobind\cite{nanobind}. To evaluate hardware-accelerated decoding, we additionally implement the FPGA pipeline described in \S~\ref{sec:hardware} and deploy it on the AMD Versal V80 accelerator platform. For the CPU version, we conducted all experiments on a host Linux system running NixOS 26.05, equipped with a 128-core Intel Xeon 6980P CPU operating at a 2.0GHz base and 3.9GHz single-core peak frequency, 1TB of main memory, and an Nvidia H200 GPU with 141GB of HBM3e vRAM.

        \myparagraph{FPGA implementation}
            We derive the FPGA pipeline from our C++ implementation via High-Level Synthesis (HLS) using AMD Vitis 2025.1 \cite{amd_vitis}. The synthesized decoding kernel is packaged as a vFPGA region within the Coyote v2 shell \cite{coyote_v2}, which provides the streaming interfaces, the virtual-address translation, and the HBM access path used by the decoding core (\S~\ref{sec:hardware}). The Versal V80 provides 32\,GB of HBM2e, connects to the host over a PCIe 5.0 $\times$8 link, and clocks the decoding kernel at the 250\,MHz provided by the Coyote shell.

\section{Evaluation}\label{sec:evaluation}

    %Organized by claims/RQs.
    %- Each result needs methodology, numbers, explanation, takeaway.
    %- Evaluate components and end-to-end behavior.

    To characterize the performance of \projectname{}, we evaluate its Logical Error Rate (LER), latency, runtime distribution, and compare results against PyMatching 2 (PM2) \cite{pymatching} (\S\ref{subsec:evaluation_ler_runtime}) utilizing the Nvidia Ising \textit{fast} pre-decoder (PD) \cite{nvidia_ising}. We then analyze scalability across varying code distances and decoder configurations (\S\ref{subsec:evaluation_scalability}). Additionally, we evaluate the hardware resource utilization of our FPGA implementation and the decoding throughput when utilizing multiple CPU cores (\S\ref{subsec:evaluation_resources_utilization}). Lastly, an ablation study isolates the impact of our design choices on decoding performance and scalability (\S\ref{subsec:evaluation_ablation}).

    %\peter{Comment from Swamit: Look at code distance 9 to 15, since we probably do not need distance 21! Swamit: I think pre-decoders produce worse LERs than a strong decoder alone. Peter: For Nvidia, this is not true. Swamit: Some results with sliding window would be nice. Swamit: How does the decoding scale with a larger number of rounds, not just d+1: Idea for an ablation study, where we increase the number of rounds. 
    %a people overemphasizing the cryostat
    %b ablation studies with corner cases!!! Error bars are important! Swamit: Doesn't parallel window decoding and sliding window fix all the issues?! Peter: %Let's show what impact our decoder has on the sliding window and the parallel window.
    %}
            \begin{figure*}[t]
    \centering
    % 2x1 Vertical stack
    \hspace{-3em}
    \begin{minipage}[b]{0.49\linewidth}
        \centering

        \includegraphics[width=.54\linewidth]{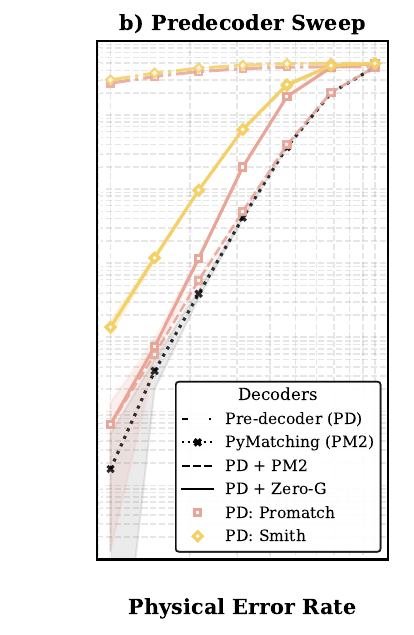}
        % Swap images
        \hspace{-25em}
        \includegraphics[width=.54\linewidth]{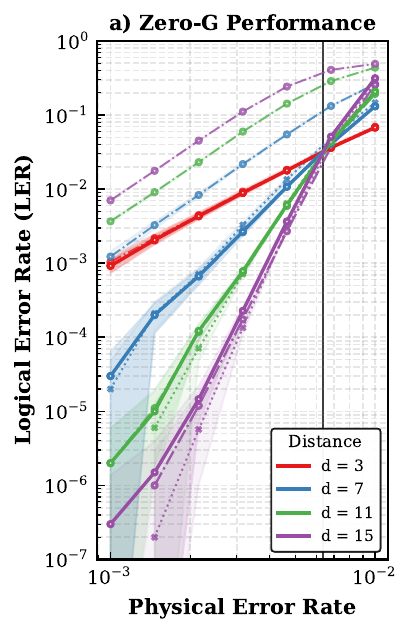}

    \end{minipage}
    \hspace{-1em}
    \raisebox{.5em}{%
        \begin{minipage}[b]{0.54\linewidth}
            \centering
            % Top Row
            \includegraphics[width=0.5\linewidth]{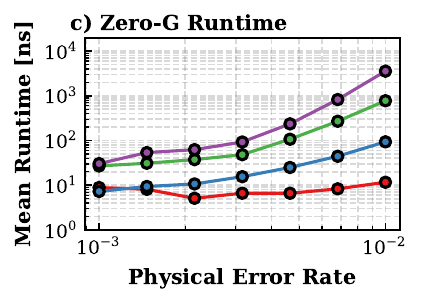} 
            \hspace{-1em}
            \includegraphics[width=0.5\linewidth]{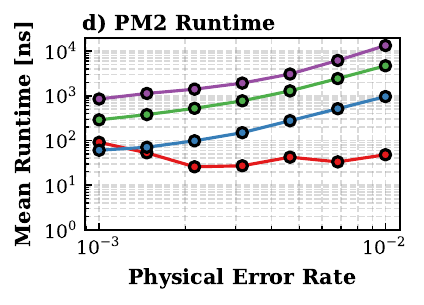} 
            
            % Bottom Row
            \vspace{.4em}
            \includegraphics[width=0.5\linewidth]{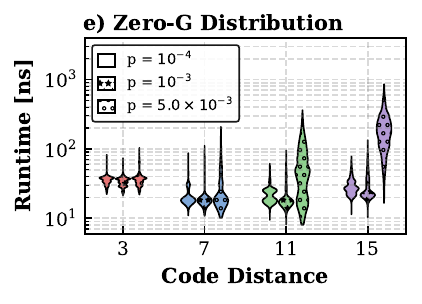} 
            \hspace{-1.3em}
            \includegraphics[width=0.5\linewidth]{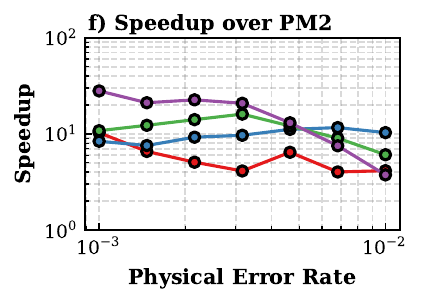}

            %\includegraphics[width=0.99\linewidth]{figures/experiments/runtime/zerog_runtime_legend.pdf}

            %\vspace{-.5em}
            %\includegraphics[width=0.95\linewidth]{figures/experiments/runtime/zerog_runtime_legend.pdf}
            
            %\vspace{-.5em}
            %\includegraphics[width=0.55\linewidth]{figures/experiments/runtime/zerog_runtime_distribution_legend.pdf}
        \end{minipage}%
    }

    \vspace{-5pt}
    \caption{
        Logical error rate, runtime, and speedup comparison for \projectname{}, PyMatching2 (PM2), and the Pre-Decoder (PD) under a surface code memory experiment.
        \emph{
        \textbf{(a)} Logical vs.\ physical error rate using Nvidia-Ising pre-decoder for multiple code distances for $k=2$. The vertical line marks the noise threshold \cite{threshold}.
        \textbf{(b)} Performance for Promatch and Smith pre-decoder for $d = 11$ and $k=100$.
        \textbf{(c)-(d)} Mean standalone decoding runtimes.
        \textbf{(e)-(f)} 99th-percentile runtime distribution and relative speedup of \projectname{} over PM2.
        }
    }
    \label{fig:experiments_performance_full}
    \vspace{-1.5em}
\end{figure*}

                        \begin{figure}[t!]
    \centering
    
    \includegraphics[width=.99\linewidth]{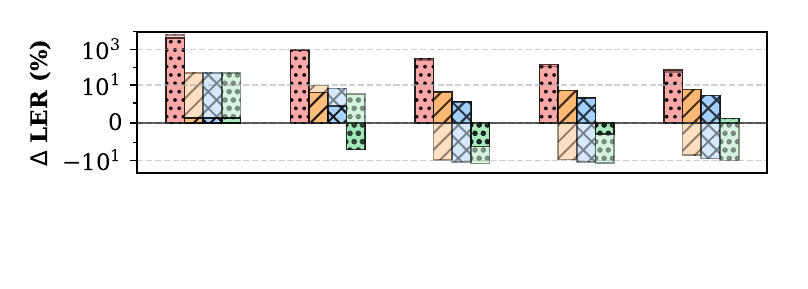}
    
    \vspace{-3.2em}
    \includegraphics[width=.99\linewidth]{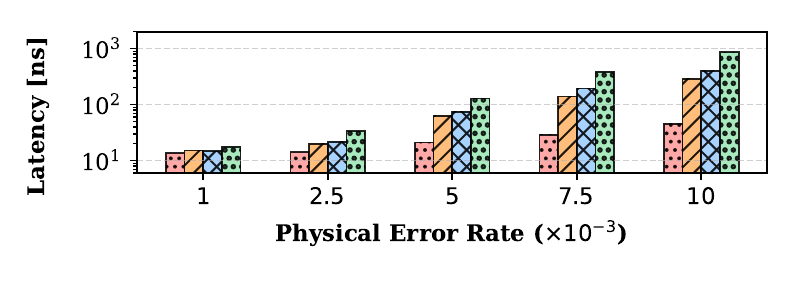}

    \vspace{-1.25em}
    \includegraphics[width=.75\linewidth]{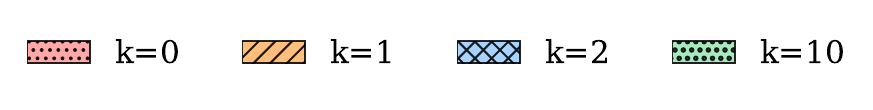}

    \vspace{-1em}
    \caption{
        Effect of \projectname{} configuration on \textbf{(a)} relative LER vs.\ PM2 and \textbf{(b)} runtime, at $d=9$.
        \emph{Solid and stacked bars show $\Delta\text{LER}$ relative to PD + PM2 and standalone PM2, respectively.}
        }
    \label{fig:experiments_configurability}
    \vspace{1pt}
    
\end{figure}

    \subsection{Accuracy and Latency}\label{subsec:evaluation_ler_runtime}

        \myparagraph{Logical error rate}
            A good decoder must be both fast and accurate, and therefore, we begin by analyzing the logical error rate (LER). Specifically, our goal is to achieve a competitive LER compared to PyMatching2.
            Fig. \ref{fig:experiments_performance_full}\textbf{a)} and \textbf{b)} visualize the LER achieved by utilizing different pre-decoders (PD) when paired with either \projectname{} or PyMatching2 (PM2), alongside standalone PD and PM2 decoding.

            %The hybrid PD + PM2 pipeline performs identically, up to statistical variance, to the standalone PM2 decoder for the Nvidia-Ising or Promatch. In contrast, a standalone PD is unable to provide meaningful decoding performance.
            %We use a naive \projectname{} configuration ($k=2$) that achieves a realistic LER-runtime trade-off at low code distances, though a larger $k$ is needed to match PM2 as $d$ grows. Overall, \projectname{} achieves LERs similar to PM2, increasing it by 2\% on average and by 40\% in the worst case ($d=15$, p=$10^{-3}$). 
            %The choice of decoder also shifts the code threshold: PM2 reaches 0.67\%, while \projectname{} reaches 0.63\%, corresponding to a worst-case LER increase of $1.58\times$ at $d=15$ following empirical power-law scaling \cite{power_law}.
            % TODO: Add short description of the second figure b)
            The hybrid PD + PM2 pipeline matches standalone PM2 for Nvidia-Ising or Promatch, whereas a standalone PD alone gives poor decoding performance. For Nvidia-Ising, PD~+~\projectname{}'s LER diverges slightly from PD~+~PM2 at low physical error rates, whereas for Promatch and Smith, it instead converges to PD~+~PM2, at the cost of more \projectname{} iterations ($k=100$). Since these iterations parallelize on FPGA, this convergence adds no runtime cost. Our naive \projectname{} configuration ($k=2$) gives a realistic LER-runtime trade-off at low distances. Overall, \projectname{} tracks PM2's LER closely, increasing it by 2\% on average and 40\% in the worst case ($d=15$, $p=10^{-3}$).

            The decoder choice also shifts the code threshold: PM2 reaches $0.67\%$ versus \projectname{}'s $0.63\%$, a worst-case LER increase of $1.58\times$ at $d=15$ under empirical power-law scaling \cite{power_law}.

            \statementLeftAccent{Takeaway}{green}{\projectname{} achieves near optimal logical error rates compared to PM2, increasing the LER by 0.1\% on average.}

        \myparagraph{Latency}
            While achieving a competitive LER is essential, real-time decoding additionally demands low decoding latency. \projectname{} offsets its slight LER degradation with substantially lower decoding latency, averaging $100\text{ ns}$ across code distances and physical error rates, as demonstrated in Fig. \ref{fig:experiments_performance_full}\textbf{c)}. 
            Concretely, at $d=9$ runtime grows from $10\text{ ns}$ to $200\text{ ns}$ as the physical error rate varies, while at $d=15$ it grows far more steeply, from $30\text{ ns}$ to $4\mu\text{s}$ ($20\times$ vs.\ $130\times$ overhead). This stems from the higher density of the sparse syndrome at larger $d$, resulting in a quadratic increase in runtime under our algorithm (\S~\ref{sec:design}). At realistic error rates $p=5\times10^{-3}$, $d=15$, we achieve 99th-percentile latencies of $200\text{ ns}$. We observe a critical error rate, coinciding with $p_{\text{training}} = 5\times10^{-3}$ reflecting the rate the pre-decoder was trained on, below which runtime scales linearly and above which it becomes quadratic.
            PM2 averages $800\text{ ns}$ under identical conditions (Fig.~\ref{fig:experiments_performance_full}\textbf{d)}), but only by relying on batch decoding across multiple syndromes, whereas single-shot decoding incurs higher latency.

            \statementLeftAccent{Takeaway}{green}{\projectname{} achieves worst case average latency of 350ns for code distances $d \leq 15$ and error rates $p \leq 5\cdot10^{-3}$.}

        \myparagraph{Speedup}
            To investigate how effectively \projectname{} performs decoding in the sparse syndrome regime, we evaluate the relative decoding latency over PM2 in Fig. \ref{fig:experiments_performance_full}\textbf{f)}. For small code distances ($d \leq 9$), \projectname{} yields an approximate $10\times$ speedup, which scales up to $30\times$ for larger code distances. Below the identified critical threshold ($p < p_{\text{training}}$), this performance advantage remains constant relative to PM2. While this speedup reduces linearly for physical error rates above $p_{\text{training}}$, \projectname{} consistently outperforms PM2.
        
            \statementLeftAccent{Takeaway}{green}{\projectname{} achieves an average speedup of $10 \times$ compared to PM2, with a maximum of $30 \times$ for $d=15$.}

        \myparagraph{Latency distribution}
            While low average latency is desirable, real-time decoding requires low \textit{worst-case} latency ~\cite{decoding_latency_influence, backlog}. What matters is the full latency \textit{distribution}, including its tail, not just average latencies \cite{decoding_latency_influence}.
            
            Fig.~\ref{fig:experiments_performance_full}\textbf{(e)} characterizes \projectname{}'s latency distribution across $p \in \{10^{-4}, 10^{-3}, 10^{-2}\}$ and $d \in \{3, 7, 11, 15\}$. For $p \in \{10^{-4}, 10^{-3}\}$, distributions are similar across the evaluated code distances and mostly fall between $30\text{ ns}$ and $130\text{ ns}$.
            At $p = 5\cdot10^{-3}$, the distribution broadens significantly for $d > 9$, reaching latencies up to $800\text{ ns}$, an effect we attribute to the pre-decoder's receptive field being optimal for $d \leq 9$, beyond which its fixed spatial window no longer fully covers all error patterns \cite{nvidia_ising}.
            Most of the distribution remains sub-$100\text{ ns}$, but rare, complex syndromes produce thin, long tails that grow with code distance, dictating worst-case latencies.

            \statementLeftAccent{Takeaway}{green}{\projectname{} achieves constant worst-case single-shot latencies of 130ns for $p < 5\cdot10^{-3}$ across all code distances.}

    \subsection{Scalability}\label{subsec:evaluation_scalability}
        Having evaluated \projectname{} under its default configuration, we next investigate how different configurations affect performance and runtime. We evaluate four configurations, $k \in \{0, 1, 2, 10\}$, ranging from a \textit{fast} fallback ($k=0$, lowest-latency) through intermediate \textit{tradeoff} points ($k=1$, $k=2$) to an \textit{accurate} configuration ($k=10$, prioritizing LER, suitable for higher-coherence platforms e.g., neutral atoms).

        Fig.~\ref{fig:experiments_configurability} compares the different configurations on LER and runtime. Latency increases roughly linearly with physical error rate across all configurations. Going from \textit{fast} to \textit{accurate} increases runtime by $20\times$ at $p=10^{-3}$ but only $3\times$ at $p=10^{-2}$, following the scalability of increasing $k$ as for varying syndrome densities (\S~\ref{subsec:impl_rama}). In return, relative LER improves substantially, though gains plateau well before runtime becomes infeasible. Thus, \projectname{} breaks the usual all-or-nothing runtime–LER trade-off common to MWPM decoders, remaining configurable while staying below $500\text{ ns}$.
        In (a), \projectname{} even outperforms PM2 and PM2+PD by up to 10\% relative LER, likely due to floating-point-to-integer rounding in PM2 that causes multiple matchings to tie in weight. However, this requires further investigation.

        \statementLeftAccent{Takeaway}{green}{\projectname{} achieves optimal LER with latencies below $300\text{ ns}$, offering configurability without sacrificing scalability.}

    \subsection{Resource Utilization}\label{subsec:evaluation_resources_utilization}
        Having established \projectname{}'s LER, latency, and scalability properties, we now turn to system-level deployment, evaluating CPU scalability and FPGA resource utilization across two target platforms: CPUs (standalone or embedded) and dedicated FPGAs.

\begin{figure}[t!]
    \hspace{-1.5em}
    \centering
    \includegraphics[width=.57\linewidth]{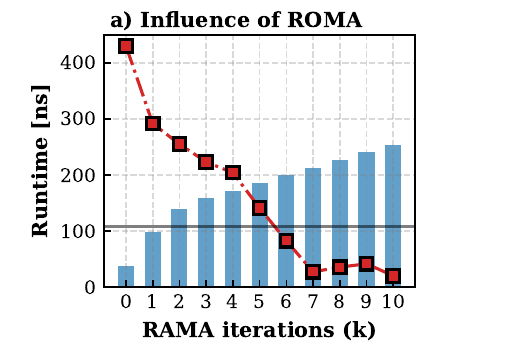}
    \hspace{-5.6em}
    \includegraphics[width=.57\linewidth]{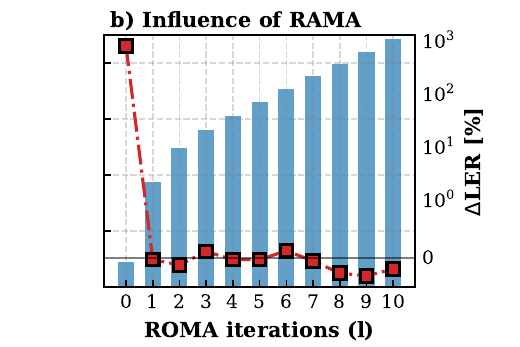}
    
    \vspace{-.5em}
    \includegraphics[width=.5\linewidth]{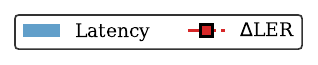}

    \vspace{-.5em}
    \caption{
        Impact of the number of \textbf{(a)} k-RAMA and \textbf{(b)} l-ROMA iterations on runtime and LER for $d=11$ and $p = 5 \cdot 10^{-3}$. \emph{\textbf{(a)} Fixed l=1, and \textbf{(b)} fixed k = 5.}
    }
    \label{fig:ablation_ler_runtime}
    \vspace{1pt}
    
\end{figure}

        \myparagraph{CPU scalability}            
            A naive implementation utilizes a single CPU core per decoding pipeline. However, it is possible to multiplex multiple decoding instances onto a single core. At $d=15$, $p=10^{-3}$, \projectname{} alone achieves $200\text{ ns}$ latency, so a single core could support up to five concurrent instances under $t_{budget}=1\mu s$, i.e.\ up to $640$ logical qubits on our $128$-core server. However, this does not account for pre-decoder latency, which in practice runs on the same budget and would reduce the achievable multiplexing factor.
    
        \myparagraph{FPGA utilization}
            Table~\ref{tab:fpga_utilization} shows post-route utilization on the Versal V80 at $d=15$ for both matrix placements of \S~\ref{sec:hardware}. The latency-optimized instance keeps only boundary records in BRAM, occupying $2.4\%$ of the LUTs, $26$ BRAM tiles, and no URAM, and the routed image closes timing at $250$\,MHz with zero routing errors. The density-optimized placement keeps the full matrices in HBM, shared read-only among instances, at a similarly small per-instance footprint. Since additional instances add only compute logic, density is bounded by the HBM pseudo-channels, supporting up to $32$ concurrent instances per V80.

        \myparagraph{FPGA latency}
            We measure on-device kernel latency at $250$\,MHz over $1{,}024$-shot batches with $k=2$ and $l=1$ on synthetic sparse residuals. With five active detectors on average, the latency-optimized pipeline decodes a shot in $0.93\,\mu\text{s}$ on average with a p99 of $1.45\,\mu\text{s}$, and when every shot carries the maximum of ten active detectors, latency averages $1.46\,\mu\text{s}$ with an observed maximum of $1.49\,\mu\text{s}$. The narrow spread confirms that latency depends only on syndrome complexity, since every stage runs a bounded number of cycles per active detector. Active-subgraph streaming reduces average kernel cycles by $16.9\times$ over the HBM design on the same workload, and a single instance sustains close to the $10^6$ syndromes/s that superconducting platforms require.

        \myparagraph{Hardware accuracy}
            The hardware pipeline computes with integer weights and permutations from a deterministic PRNG (\S~\ref{sec:hardware}). A software model of these hardware semantics validates every decoded output, and all $20{,}480$ measured shots match it exactly, confirming that command framing, matching, and observable construction operate correctly on hardware.

\begin{figure}[t!]
    \centering
    
    \includegraphics[width=.98\linewidth]{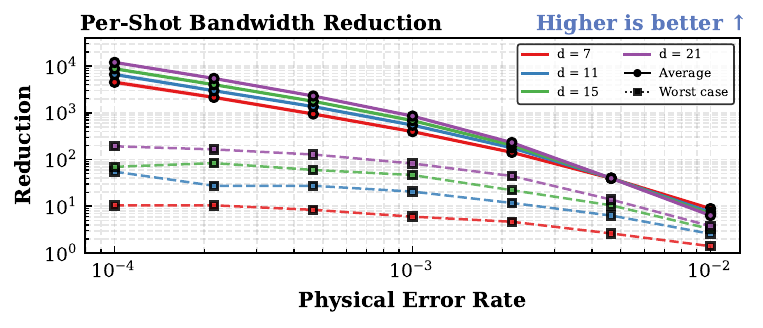}

    \vspace{-1em}
    \caption{
        Average and worst-case per-shot bandwidth savings for communication between the NVIDIA-Ising pre-decoder and the \projectname{} strong decoder, achieved by passing active detectors instead of the full syndrome. \emph{Rather than transmitting the entire syndrome, we send only the IDs of active detectors to \projectname{}.}
        }
    \label{fig:experiments_bandwidth_saving}
    \vspace{1pt}
    
\end{figure}

        \myparagraph{Bandwidth reduction}
            As shown in Fig.~\ref{fig:experiments_bandwidth_saving}, by transmitting active detector IDs instead of full syndromes (\S~\ref{subsec:fpga_architecture}) we achieve substantial per-shot bandwidth savings, averaging $4\cdot10^3\times$ ($d=7$) to over $10^4\times$ ($d=21$) and remaining above $10\times$ even in the worst case. This reduction grows with code distance, since larger codes have proportionally sparser syndromes, thereby easing both the worst-case bandwidth and thermal constraints that otherwise bottleneck cryogenic decoders \cite{pinball, clique, afs_decoder, instruction_bandwidth}, and making \projectname{} well-suited for such deployments.

        \statementLeftAccent{Takeaway}{green}{\projectname{} supports up to 640 logical qubits on a 128-core CPU. On a single V80, a latency-optimized instance decodes typical $d=15$ residuals in $0.93\,\mu\text{s}$, and the density-optimized deployment scales to 32 instances.}

    \subsection{Ablation Study}\label{subsec:evaluation_ablation}
        We now isolate the individual factors driving \projectname{}'s performance, namely the number of RAMA/ROMA iterations, syndrome complexity, window scaling, and end-to-end pipeline latency.
% Real anchors (rose, V80/Vivado 2025.1):
% zerog-d15-v80-rama-triangular-changed (deployed, boundary-only streaming),
% as-built-v80-implementation.md, 2026-07-29:
%   user region synth: LUT 62,106 (2.41)  Reg 54,998 (1.07)  BRAM 26 (0.70)  URAM 0
%   routed cyt_top:    LUT 125,859 (4.89) Reg 179,893 (3.49) BRAM 58 (1.55)  URAM 0
%   (timing closed: WNS +0.001, zero routing errors)
% zerog-d15-v80-rama (HBM), routed cyt_top (reports/shell_utilization.rpt):
%   LUT 127,558 (4.96)  Reg 248,594 (4.83)  BRAM 121.5 (3.25)  URAM 4 (0.21)
%   user region synth (reports/config_0/user_synthed_c0_0.rpt):
%   LUT 23,876 (0.93)  Reg 43,476 (0.84)  BRAM 83.5 (2.23)  URAM 4 (0.21)
% Shell rows are derived as (routed cyt_top - decoder region), so each block
% sums to its measured routed total.
\begin{table}[t]
    \centering
    \caption{\projectname{} post-route resource utilization on the AMD Versal V80 for $d=15$. \emph{The latency-optimized placement keeps only boundary records in BRAM and receives active-pair records with each shot, whereas the density-optimized placement keeps the full matrices in HBM shared across decoder instances.}}
    \label{tab:fpga_utilization}
    \setlength{\tabcolsep}{3pt}
    \resizebox{\columnwidth}{!}{%
    \begin{tabular}{|l|rr|rr|rr|rr|}
    \hline
    \textbf{Entity} & \multicolumn{2}{c|}{\textbf{LUT (\%)}} & \multicolumn{2}{c|}{\textbf{Register (\%)}} & \multicolumn{2}{c|}{\textbf{BRAM (\%)}} & \multicolumn{2}{c|}{\textbf{URAM (\%)}} \\ \hline
    Versal V80                 & 2,574,208 & (100)  & 5,148,416 & (100)  & 3,741 & (100)  & 1,925 & (100)   \\ \hline
    \multicolumn{9}{|l|}{\textit{Latency-optimized placement (active-subgraph streaming)}} \\ \hline
    Shell                      & 63,753    & (2.48) & 124,895   & (2.43) & 32.0  & (0.86) & 0     & (0.00)  \\
    Decoder                    & 62,106    & (2.41) & 54,998    & (1.07) & 26.0  & (0.70) & 0     & (0.00)  \\ \hline
    \multicolumn{9}{|l|}{\textit{Density-optimized placement (HBM, shared)}} \\ \hline
    Shell                      & 103,682   & (4.03) & 205,118   & (3.98) & 38.0  & (1.02) & 0     & (0.00)  \\
    Decoder                    & 23,876    & (0.93) & 43,476    & (0.84) & 83.5  & (2.23) & 4     & (0.21)  \\ \hline
    \end{tabular}%
    }
\end{table}

        \myparagraph{RAMA/ROMA steps}
            Fig.~\ref{fig:ablation_ler_runtime} shows runtime and relative LER vs.\ PM2 as a function of RAMA and ROMA rounds at fixed code distance $d=11$. While runtime grows linearly with the number of rounds, $\Delta$LER improves quickly at first and then plateaus around $k \approx 5$ and $l \approx 1$. Beyond this point, additional rounds yield diminishing returns.

        %\myparagraph{Syndrome complexity}
        %    To isolate the impact of syndrome complexity from problem size, we measure decoding latency across code distances $d= [5, 11]$ for fixed Hamming weights (Fig.~\ref{fig:ablation_hw_runtime}\textbf{a)} and \textbf{b)}). As shown, latency is determined solely by syndrome complexity rather than code distance, since latencies match for both $d$ with equal Hamming weights. Thus, any reduction in syndrome complexity by a pre-decoder directly improves \projectname{}'s runtime. Paired with a deterministic pre-decoder such as ProMatch \cite{promatch}, which caps syndrome complexity at $10$, \projectname{} can guarantee strictly deterministic latencies.

\begin{figure}[t!]
    \centering
    \includegraphics[width=.53\linewidth]{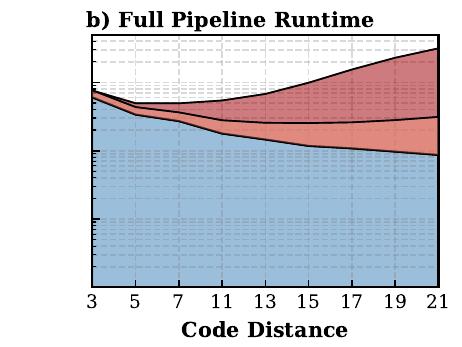}
    \hspace{-25em}
    \includegraphics[width=.53\linewidth]{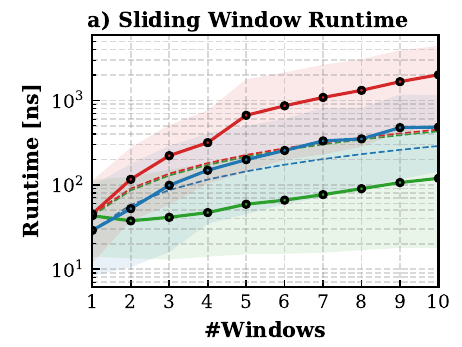}

    \vspace{-.5em}
    \includegraphics[width=.8\linewidth]{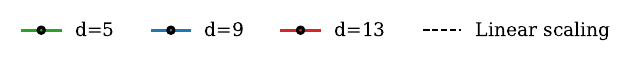}
    
    \vspace{-.85em}
    \includegraphics[width=.9\linewidth]{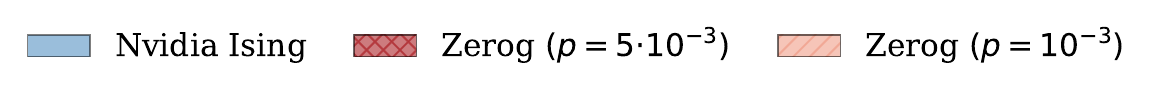}
    
    \vspace{-.5em}
    \caption{
        \textbf{(a)} Decoding runtime vs.\ number of windows decoded for at $p = 10^{-3}$.
        \textbf{(b)} Full Nvidia-Ising + \projectname{} pipeline latency. %for $p \in \{10^{-3}, 5\cdot10^{-3}\}$.
    }
    \label{fig:ablation_window_scaling}
    \vspace{1pt}
    
\end{figure}

        \myparagraph{Sliding window decoding}
            Beyond scaling with syndrome complexity, we evaluate scaling with the number of syndrome-extraction rounds, simulating sliding-window or parallel-window deployment. As shown in Fig.~\ref{fig:ablation_window_scaling}, when scaling the number of windows, where $d$ rounds equal one window, both average and worst-case decoding latency scale approximately linearly with window count. This makes \projectname{} well-suited to sliding-window and parallel-window pipelines \cite{sliding_window, parallel_window} such as NVIDIA CUDA-Q Realtime \cite{nvidia_cudaq_realtime}.

        \myparagraph{Total runtime}
            Fig.~\ref{fig:ablation_window_scaling}\textbf{(b)} shows total end-to-end pipeline runtime. At small code distances, latency is dominated by the pre-decoder, whereas as $d$ increases, \projectname{}'s share increases with syndrome complexity due to increased syndrome size. At $p=10^{-3}$, the pipeline comfortably stays under $1\mu\text{s}$ up to $d = 21$, and at $p=5\times10^{-3}$, it meets the required throughput up to $d \leq 15$.

            Note, however, that these numbers exclude CPU-to-GPU host-to-device transfer overhead and GPU kernel startup, both of which could be mitigated by migrating both the pre-decoder and \projectname{} onto a unified FPGA.%, or via direct FPGA-to-GPU transfer.

        \statementLeftAccent{Takeaway}{green}{\projectname{}'s latency scales predictably with window count, its accuracy saturates after only a few RAMA/ROMA rounds, and, accounting for pre-decoder latency, the full pipeline stays within the $1\,\mu\text{s}$ budget up to $d=15$.}

\section{Related Work} \label{sec:related_work}
    
    Real-time decoding is a central bottleneck on the path to fault-tolerant quantum computing, leading to a large and growing body of work on decoding algorithms. Real-time decoders fall broadly into three categories: greedy decoding, general decoders, and pre-decoders.
    
    \myparagraph{Greedy decoding}
        Greedy algorithms trade accuracy for lower latency by approximating, e.g., the MWPM problem.
        Vittal et al. perform a brute-force search to scale to $d=7$, with a reduced-accuracy variant reaching $d=9$ on an FPGA \cite{astrea}.
        Liao et al. extend this approach by using a Lookup-Table to support code distances up to $d=13$ \cite{wit_greedy}.
        Forlivesi et al. propose spanning tree matching on the detector graph \cite{spanning_decoder}, but it is limited to fewer than 5 triggered detectors for competitive LER and runtime.
        None of these approaches simultaneously scales to practically relevant code distances while achieving the required accuracy.

    \myparagraph{General decoders}
        General-purpose decoders aim for an optimal decoding solution and the best possible LER.
        Higgott et al. propose \textit{Sparse Blossom}, the state-of-the-art MWPM decoder\cite{pymatching, blossom}, whereas Wu et al. propose a FPGA implementation \cite{microblossom} with reduced latency. In contrast, Delfosse et al. propose union-finding decoding, an approximation to MWPM \cite{union_finding}, whereas Liyanage et al. implement it on an FPGA, reaching $d=17$, at the cost of high FPGA resources \cite{helios}. To support general qLDPC codes, Müller et al. propose a decoder that is similarly resource-intensive and tailored to FPGAs \cite{relaybp}. More recently, machine learning approaches use a neural network solver but are generally limited to $d=11$ \cite{alphaqubit2}.
        However, all existing decoders thus either fail to scale to necessary code distances or require unscalable hardware resources.
    
    \myparagraph{Pre-decoders}
        %Syndrome-modifying pre-decoders reduce the latency of strong decoders by removing easy-to-solve errors before passing syndromes on.
        Smith et al. propose a greedy syndrome-modifying approach matching nearby detectors at the cost of reduced accuracy \cite{smith_predecoding}.
        Alavisamani et al. iteratively remove simple errors until at most 10 triggered detectors remain \cite{promatch}, combining this with Astrea to reach $d=13$. Chamberland et al. propose a neural network-based pre-decoder with good accuracy and constant runtime to improve the scalability of PyMatching2 \cite{nvidia_ising}. %Non-syndrome-modifying cryogenic pre-decoders reduce the workload of strong decoders but not the syndrome complexity, and thus cannot reduce worst-case decoding latency \cite{clique, pinball, arqade}.
        By relying on generic decoders for remaining errors, current pre-decoders lose much of their potential.

\section{Conclusion}

    We present \projectname{}, a novel decoder explicitly designed for use with pre-decoders, enabling it to fully exploit sparse residual syndromes. Our novel stochastic approximate MWPM decoding algorithm matches PyMatching~2's logical error rate while decoding  $10\times$ faster on average, while supporting varying accuracy-latency parameterization. By separating control from the decoding data plane, we support hardware-agnostic deployment on CPUs and FPGAs. A single 128-core CPU supports 640 logical qubits, while a single Versal V80 FPGA decodes typical $d=15$ residuals in under a microsecond, scaling to 32 logical qubits.

\myparagraph{Artifact} \projectname{} will be publicly available for the artifact evaluation, along with the entire experimental setup.

\section*{Acknowledgment}
        This work was funded by the Bavarian State Ministry of Science and the Arts as part of the Munich Quantum Valley (MQV) initiative, grant number 6090181.

%%%%%%% -- PAPER CONTENT ENDS -- %%%%%%%%

%%%%%%%%% -- BIB STYLE AND FILE -- %%%%%%%%
\bibliographystyle{IEEEtranS}
\bibliography{refs}
%%%%%%%%%%%%%%%%%%%%%%%%%%%%%%%%%%%%

\end{document}